\documentclass[a4paper]{spie} 
\usepackage[]{graphicx}
\usepackage{amssymb,amsmath, psfrag}
\usepackage{algorithm}
\usepackage{algorithmic}
\usepackage{psfrag} 

\newcommand{\cnum}{\mathbb{C}}

\newcommand{\rnum}{{\bf {R}}}

\newcommand{\curl}{\mathrm{curl}\;}
\newcommand{\divo}{\mathrm{div}\;}

\newcommand{\dd}[1]{\mathrm{d}\,#1}

\newcommand{\VField}[1]{{\bf #1}}

\def\squareforqed{\hbox{\rlap{$\sqcap$}$\sqcup$}}
\def\qed{\ifmmode\else\unskip\quad\fi\squareforqed}

\title{Advanced Finite Element Method for Nano-Resonators}
\author{Lin Zschiedrich\supit{a\,b}, Sven Burger\supit{a\,b}, 
Benjamin Kettner\supit{a}, and Frank Schmidt\supit{a\,b}
\skiplinehalf
\supit{a} Zuse Institute Berlin (ZIB), Takustra{\ss}e 7, D-14195 Berlin, Germany \\
\supit{b} JCMwave GmbH, Haarer Stra{\ss}e 14a, D-85640 Putzbrunn, Germany
}

\authorinfo{Further author information: (Send correspondence to Lin Zschiedrich) \\
E-mail: zschiedrich@zib.de \\
URL: http://www.zib.de/nano-optics/ 
}

\begin{document} 
\maketitle 

\noindent
Copyright 2006  Society of Photo-Optical Instrumentation Engineers.\\
This paper will be published in Proc.~SPIE {\bf 6115} (2006), 
({\it Physics and Simulation of Optoelectronic Devices XIV}).
and is made available 
as an electronic preprint with permission of SPIE. 
One print or electronic copy may be made for personal use only. 
Systematic or multiple reproduction, distribution to multiple 
locations via electronic or other means, duplication of any 
material in this paper for a fee or for commercial purposes, 
or modification of the content of the paper are prohibited.

\begin{abstract}
Miniaturized optical resonators with spatial dimensions of 
the order of the wavelength of the trapped light offer 
prospects for a variety of new applications like  
quantum processing or construction of meta-materials. Light propagation
in these structures is modelled by Maxwell's equations.
For a deeper numerical analysis one may compute the scattered
field when the structure is illuminated or one may compute
the resonances of the structure. We therefore address in this 
paper the electromagnetic scattering problem as well as
the computation of resonances in an open system. 
For the simulation efficient and reliable numerical methods
are required which cope with the infinite domain. 
We use transparent boundary conditions based
on the Perfectly Matched Layer Method (PML) combined with a
novel adaptive strategy to determine optimal discretization
parameters like the thickness of the sponge layer or the mesh width.
Further a novel iterative solver for time-harmonic Maxwell's
equations is presented.   
\keywords{Nano-Optics, Meta-Materials, Resonances, Scattering, Finite-Element-Method, PML}
\end{abstract}
\section{INTRODUCTION}
\label{sect:intro} 
With the advances in nanostructure physics it has become possible 
to construct light resonators on a lengthscale equal to or even
smaller than optical wavelengths~\cite{Linden2004a,Enkrich2005a}.
These nanostructures are large on the atomic scale, therefore they 
can be of complex geometry and they may possess properties not 
occuring in nature, like an effective negative index of 
refraction~\cite{Veselago1968a}
which allows in principle to overcome limits in the resolution of 
optical imaging systems~\cite{Pendry2000a}. 

The numerical simulation of light fields in 
such structures is a field of ongoing research. 
In this paper we report on finite element methods for the efficient 
computation of resonances and light propagation in arbitrarily 
shaped structures embedded in simply structured, infinite domains.
Section \ref{geometric_section} introduces our concept of  discretizing 
exterior infinite domains.
Section \ref{scattering_section} recapitulates a formulation of Maxwell's equations for 
time-harmonic scattering problems.
Section \ref{pml_section} introduces an adaptive method for the efficient discretization 
of the exterior domain based on the PML method introduced by Berenger~\cite{Berenger94}.
Section \ref{weak_section} shows the weak formulation of Maxwell's equations which 
is needed for the finite-element method. 
In Section \ref{td_section} we shortly introduce a new preconditioner for the numerical
solution of indefinite time-harmonic Maxwell's equations. 
Finally, in Sections \ref{srr_section} and \ref{pyramid_section} we test our algorithms on 
nano-optical real world problems: the computation of resonances and 
scattering in arrays of split-ring resonators and in isolated pyramidal 
nano-resonators.

\section{GEOMETRIC CONFIGURATION}
\label{geometric_section}
In this section we explain how to specify an infinite geometry such that it fits
well to the finite element method (FEM). The geometry is split into a bounded interior domain
 $\Omega_{\mathrm{int}}$ and an unbounded exterior domain $\Omega_{\mathrm{ext}}.$ 
The interior domain may contain nearby arbitrary shaped structures such as spheres or thin layers. 
The geometry in the exterior domain is more restricted. However, 
the construction we propose is general enough to deal with typical geometries of optical devices.

\begin{figure}
  \begin{center}
    \psfrag{F1}[lc][lc][1.2][0]{$F_{1}$}
    \psfrag{F2}[lc][lc][1.2][0]{$F_{2}$}
    \psfrag{QF1}[lc][lc][1.2][0]{$Q_{F_{1}}$}
    \psfrag{xi}[lc][lc][1.2][0]{$\xi$}
    \psfrag{eta1}[lc][lc][1.2][0]{$\eta_{1}$}
    \psfrag{eta2}[lc][lc][1.2][0]{$\eta_{2}$}

    \includegraphics[width = 0.6\textwidth]{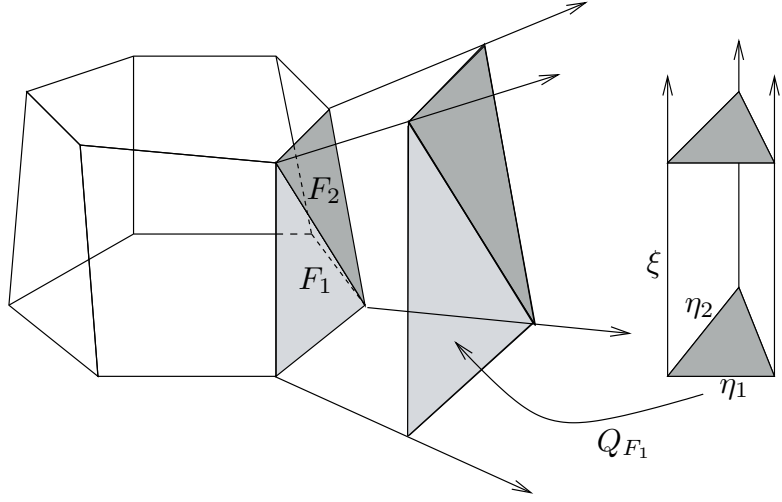}
    \caption{Infinite domain. The interior domain (left) may contain
nearby arbitrarily shaped objects. The exterior domain consists of prisms attached to
triangular boundary faces of the interior domain. Each prism is the image of the unit prism (right)
under a bilinear mapping such that the triangles with $\xi=const$ are
mapped to parallel triangles. 
For each infinite prism we assume constant material parameters.}
    \label{Fig.ScetchCompDomain}
  \end{center}
\end{figure}

We assume that the boundary $\Gamma$ of the interior domain consists 
of triangles.
A boundary triangle $F \subset \Gamma$ is called {\em transparent} 
if $F \subset \partial \Omega_{\mathrm{ext}}.$ Further
we introduce the unit prism 
$P_{\mathrm{u}}=\{(\eta_{1}, \eta_{2}, \xi) \in \rnum^{3}\;:\; \eta_{1}, \eta_{2}, \xi \geq 0, 
\eta_{1}+\eta_{2}\leq 1\}.$
An exterior domain is admissible if the following conditions are
satisfied. 
For each boundary triangle $F$ there exists a bilinear 
one-to-one mapping $Q_{F}$ from the unit prism into the exterior domain
$\Omega_{\mathrm{ext}}$ such that each triangle 
$T_{\rho}=\{(\eta_{1}, \eta_{2}, \xi) \subset P_{\mathrm{u}}\;:\;\xi=\rho\}$ is mapped
onto a triangle parallel to the face and such that the bottom triangle of the unit
patch is mapped onto the corresponding face, $Q_{F}T_{0}=F,$ cf. Figure~\ref{Fig.ScetchCompDomain}.
The image of $Q_{F}$ is denoted by $P_{F}.$ 
Hence we attach the infinite prism $P_{F}$ to the transparent face $F.$ It must hold true 
that $\Omega_{\mathrm{ext}}=\cup_{F}P_{F}.$ Further we demand the following matching 
condition. If $Q_{F}(\eta_{1}, \eta_{2}, 0) = Q_{F'}(\eta'_{1}, \eta'_{2}, 0)$, that is
$F$ and $F'$ have a common point, then  
$Q_{F}(\eta_{1}, \eta_{2}, \xi) = Q_{F'}(\eta'_{1}, \eta'_{2}, \xi)$ for all $\xi \in \rnum_{+}.$

The surface $S_{\rho}= \cup_{F}Q_{F}T_{\rho}$ looks like a ``stretched'' transparent boundary of
the interior domain. Hence the coordinate $\xi$ is chosen consistently for all prism such that
it serves as a generalized distance variable. 
This is essential for the pole condition concept developed
by Frank Schmidt~\cite{Schmidt2002a}. 
For later purposes we introduce 
the truncated unit prism $P_{\rho}=\{(\eta_{1}, \eta_{2}, \xi) \in \rnum^{3}\;:\; \eta_{1}, \eta_{2} \geq 0, 
\eta_{1}+\eta_{2}\leq 1, 0\leq \xi \leq \rho\}$ and the truncated
exterior domain $\Omega_{\rho}=\cup_{F}Q_{F}P_{\rho}.$

If there exist triangles on $\Gamma$ which are not transparent, then either 
boundary conditions must be imposed on them, or they must be identified with 
other periodic triangles (e.g., when $\Omega$ is a cell of a periodic structure). However
for simplicity we assume that $\Omega_{\mathrm{int}} \cup \Omega_{\mathrm{ext}}=\rnum^{3}$ 
in rest of the paper.

\section{SCATTERING PROBLEMS}
\label{scattering_section}
Monochromatic light propagation in an optical material is modelled by the time-harmonic
Maxwell's equations
\begin{subequations}
  \label{eq.THMaxSystem}
  \begin{eqnarray}
    \curl \mu^{-1}\left(\vec{x} \right) \curl \VField{E}\left(\vec{x}\right) -
    \omega^{2} \varepsilon \left(\vec{x}\right)\VField{E}\left(\vec{x}\right) & = & 0,
    \label{eq.THMax}
    \\ \divo \varepsilon \left(\vec{x}\right) \VField{E}\left(\vec{x}\right) & = & 0,
    \label{eq.DivCond}
  \end{eqnarray}
\end{subequations}
which may be derived from Maxwell's equations when assuming a
time dependency of the electric field as $\VField{E}(\vec{x}, t) =
\VField{E} \left(\vec{x} \right) \exp(-i \omega t)$ with angular frequency
$\omega$. The dielectric tensor $\varepsilon$ and the permeability tensor $\mu$ are
$L^{\infty}$ functions of the spatial variable $\vec{x}=(x_{1}, x_{2}, x_{3})$. In addition we assume that the tensors $\varepsilon$ and $\mu$ 
 are constant on each infinite prism as defined in the previous section.
 For simplicity assume that the dielectric and the permeability
tensors are isotropic so they may be treated as scalar valued functions. Recall that any solution to~\eqref{eq.THMax} with $\omega \neq 0$ also meets the
divergence condition~\eqref{eq.DivCond}. 

A scattering problem may be defined as follows: Given an incoming electric field
$\VField{E}_{\mathrm{inc}}$ satisfying the time-harmonic Maxwell's
equations~\eqref{eq.THMaxSystem} for a fixed angular frequency 
$\omega$ in the exterior domain, compute the total
electric field $\VField{E}$ satisfying~\eqref{eq.THMaxSystem} in 
$\Omega_{\mathrm{int}} \cup \Omega_{\mathrm{ext}}$, such
that the scattered field $\VField{E}_{\mathrm{sc}} =
\VField{E}-\VField{E}_{\mathrm{inc}}$ defined 
on $\Omega_{\mathrm{ext}}$ is {\em outward radiating}. For a precise
definition of when a field is outward radiating we refer to Schmidt~\cite{Schmidt2002a}.
Hence the scattering problem splits into an interior subproblem
for $\VField{E}_{\mathrm{int}}=\VField{E}_{|\Omega_{\mathrm{int}}} $
on $ \Omega_{\mathrm{int}}$
\begin{eqnarray*}
  \curl \mu^{-1} \curl \VField{E}_{\mathrm{int}} -
  \omega^{2} \varepsilon \VField{E}_{\mathrm{int}}
 & = & 0,
\end{eqnarray*}
and an exterior subproblem on $ \Omega_{\mathrm{ext}}$
\begin{eqnarray*}
  \curl \mu^{-1} \curl \VField{E}_{\mathrm{sc}} -
  \omega^{2} \varepsilon \VField{E}_{\mathrm{sc}} & = & 0.
\end{eqnarray*}
These subproblems are coupled by the following matching conditions:
\begin{eqnarray}
\VField{E}_{\mathrm{int}} \times \vec{n} & = & 
\left(\VField{E}_{\mathrm{inc}} + \VField{E}_{\mathrm{sc}} \right) \times \vec{n} \\
\mu^{-1}\curl \VField{E}_{\mathrm{int}} \times \vec{n} & = & 
\mu^{-1}\curl \left(\VField{E}_{\mathrm{inc}} + \VField{E}_{\mathrm{sc}} \right) \times \vec{n} 
\end{eqnarray}
on the boundary $\partial \Omega_{\mathrm{int}}.$

\section{ADAPTIVE PML METHOD}
The perfectly matched layer method was originally introduced by Berenger in 1994~\cite{Berenger94}. The idea is to discretize a complex continued field in the exterior domain which decays exponentially fast with growing distance to the interior-exterior domain coupling boundary. This way 
a truncation of the exterior domain only results in small artificial reflections. 
The exponential convergence of the method with growing thickness of the sponge layer was proven for homogeneous exterior domains by Lassas and Somersalo~\cite{Lassas:98a,Lassas:01a}. 
An alternative proof with a generalization to a certain type of inhomogeneous exterior domain is given by Hohage et al~\cite{Hohage01b}. Nevertheless as shown in our paper~\cite{Schaedle:06a} the PML method intrinsically fails for certain types of exterior domains such as layered media. This is due to a possible total reflection at material interfaces. In this case there exists a {\em critical} angle of incidence for which the resulting field in the exterior domain is neither propagating nor evanescent.
\label{pml_section}
\begin{algorithm}
  \caption{Adaptive PML method}
  \begin{algorithmic}[]
    \label{Algorithm:AdaptivePML}
    \REQUIRE $\epsilon, \sigma, h_{\mathrm{int}}, \kappa_{\mathrm{min}}$ 
    \STATE Compute $N_{\mathrm{p.w}}$ and $\xi_{\mathrm{max}}$ depending on $h_{\mathrm{int}}$ and 
    finite element order 
    \WHILE{(not converged)}
    \STATE 
    $\xi_{0}=0.0; \xi_{1}=h_{\mathrm{int}};N=1;$
    \WHILE {($-\ln(\epsilon)/(\xi_{N}\sigma)<\kappa_{\mathrm{min}}$)}
    \STATE
    $\xi_{N+1} = \xi_{N}+
    \max\{h_{\mathrm{int}},\; 2\pi \sigma \xi_{N} /(-\ln(\epsilon))/N_{\mathrm{p.w}} \}.
    $
    \IF {($\xi_{N+1} > 1/\epsilon$)}
    \STATE 
    \bf{break}
    \ELSE
    \STATE
    $N=N+1$
    \ENDIF
    \ENDWHILE
    \STATE Compute solution $u$ with PML discretization 
    $\{\xi_{0}, \xi_{1}, \dots, \xi_{N}\}$
    \IF {$\|u(\cdot, \xi_{N})\| \leq \epsilon \|u(\cdot)\|$}
    \STATE
    converged
    \ELSIF {$\xi_N>\xi_{\mathrm{max}}$}
    \STATE 
      break
    \ELSE
    \STATE
    $\kappa_{\mathrm{min}} = \kappa_{\mathrm{min}}/2$
    \ENDIF
    \ENDWHILE
  \end{algorithmic}
\end{algorithm}
Here we show that it is possible to overcome these difficulties when using an adaptive method
for the discretization of the exterior domain problem.
We assume the following expansion of the scattered field in the exterior domain
\begin{equation}
  \label{Eqn:PMLExpansion}
  \VField{E}_{\mathrm{sc}}\left(\eta_{1}, \eta_{2}, \xi \right) \sim \int \VField{c}(\eta_{1}, \eta_{2}, \alpha) 
  e^{ik_{\xi}(\alpha)\xi}\,\dd{\alpha}
\end{equation} 
with $\Re k_{\xi}(\alpha) \geq 0, \Im k_{\xi}(\alpha)\geq 0$ and a bounded function
$\VField{c}(\eta_{1}, \eta_{2}, \alpha).$ Hence $\VField{E}_{\mathrm{sc}}$  is a superposition of outgoing or 
evanescent waves in $\xi$ direction. In our notation 
we have assumed that there exists a {\em global} $(\eta_{1}, \eta_{2}, \xi)$-coordinate system
for the exterior domain. But in the following only the global meaning of the $\xi$ coordinate as explained in Section~\ref{geometric_section} will be used, so $\eta_{1}$ and $\eta_{2}$ may also be considered as coordinates of a local chart for a subdomain of $\partial \Omega_{\mathrm{int}}.$
For $\gamma = 1+i\sigma$ the complex continuation, 
$\xi \mapsto \gamma \xi,$ 
$\VField{E}_{\mathrm{sc}, \gamma}(\cdot, \cdot, \xi) = 
\VField{E}_{\mathrm{sc}}(\cdot, \cdot, \gamma \xi)$ gives
\begin{equation}
\label{Eqn:ExpDecay}
|\VField{E}_{\mathrm{sc}, \gamma} | \leq e^{-\kappa x_{2}} C,
\end{equation}
with $\kappa = \min_{\alpha}\{\Im{k(\alpha)}, \sigma \Re{k(\alpha)} \}$. Therefore
$\VField{E}_{\mathrm{sc}, \gamma}(\cdot, \cdot, \xi)$ decays exponentially fast with
growing generalized distance $\xi$ to the coupling boundary.
The idea
is to restrict the complex continuation of the exterior domain problem to a truncated domain
$\Omega_{\rho}$ and to impose a zero Neumann boundary
condition at $\partial \Omega_{\rho}$. In the next section we will give a corresponding 
variational problem which can be discretized with the finite element method where we will use
a tensor product ansatz in the truncated exterior domain $\Omega_{\rho}$ based on the 
triangulation of the surface $\partial \Omega_{\mathrm{int}}$ and a 1D mesh in 
$\xi$-direction, $\{0, \xi_{1}, \xi_{2}, \dots\, \xi_{N}\}.$ In this section we present an algorithm for the automatic determination of optimal discretization points $\xi_{j}.$

As can be seen from Equation~\eqref{Eqn:ExpDecay} the PML 
method only effects the outgoing part with $\Re k_{\xi}$ strictly larger than 
zero. Field contributions with an large $\Re k_{\xi}$ component are efficiently
damped out.  Furthermore evanescent field contributions are damped out independently
of the complex continuation.  For a proper approximation of the oscillatory and
exponential behavior a discretization that is fine enough is needed to resolve the field.  In
contrast to that anomalous modes or ``near anomalous'' modes with $k_{\xi} \sim 0$
enforce the usage of a large $\rho$ but can be well approximated with a relatively
coarse discretization in $\xi$. Such ``near anomalous'' modes typically occur
in the periodic setting but may also be present for isolated structures with a layered exterior
domain \cite{Petit1980a}. 
Hence for an efficient numerical approximation of the scattered field one must
use an adaptive discretization.  It is useful to think of the complex continuation as a
high-frequency filter. With a growing distance $\xi$ to the interior coupling
boundary the higher frequency contributions are damped out so that the discretization
can be coarsened.

For a given threshold $\epsilon$ we introduce the cut-off function
\[
\kappa_{\mathrm{co}, \epsilon}(\xi) = -\ln(\epsilon)/\xi~.
\]
At $\xi'>0$ each component in the expansion~(\ref{Eqn:PMLExpansion}) with
$\kappa>\kappa_{\mathrm{co}, \epsilon}(\xi')$ is damped out by a factor smaller than the
threshold $\epsilon,$
\[
e^{-\kappa \xi'} < e^{-\kappa_{\mathrm{co}, \epsilon}(\xi') \xi} = 
e^{\ln(\epsilon)} = \epsilon.
\]
Assuming that this damping is sufficient we are allowed to select a discretization
which must only approximate the lower frequency parts with $\kappa \leq
\kappa_{\mathrm{co}, \epsilon}(\xi)$ for $\xi>\xi'.$ If we use a fixed number
$N_{\mathrm{p.w}}$ of discretization points per (generalized) wavelength $2\pi
/\kappa$ we get the following formula for the {\em a priori} determination of the
local mesh width $h({\xi}) = 2\pi \sigma
/\kappa_{\mathrm{co}, \epsilon}(\xi)/N_{\mathrm{p.w}}.$ Since $\kappa_{\mathrm{co}, \epsilon}(\xi)
\rightarrow \infty$ for $\xi \rightarrow 0$ the local mesh width is zero at $\xi=0.$
As it is not reasonable to use a finer discretization in the exterior domain than in
the interior domain we bound the local mesh width by the minimum mesh width
$h_{\mathrm{int}}$ of the interior domain discretization on the coupling boundary, 
\[
h({\xi}) = \max\{h_{\mathrm{int}},\; 2\pi \sigma /\kappa_{\mathrm{co}, 
  \epsilon}(\xi)/N_{\mathrm{p.w}} \}.
\] 
The parameters $\epsilon$ and $N_{\mathrm{p.w}}$ are also fixed accordingly to the
interior domain discretization quality. The grid $\{\xi_{0}, \xi_{1}, \xi_{2}, \dots
\}$ is recursively constructed by
\[
\xi_{n+1} = \xi_{n}+h(\xi_{n}).
\]
This way $\xi_{n}$ grows exponentially with $n.$ To truncate the grid we assume
that components in the expansion with $\kappa<\kappa_{\mathrm{min}}$
can be neglected so that the grid $\{\xi_{0}, \xi_{1}, \dots, \xi_{N}\}$ is
determined by $ \kappa_{\mathrm{co}, \epsilon}(\xi_{N}) <
\kappa_{\mathrm{min}}\leq\kappa_{\mathrm{co}, \epsilon}(\xi_{N-1}).$ 
\\
As an {\em a posteriori} control we check if the field is indeed sufficiently
damped out at $\xi_{N}$, $\|u(\cdot, \xi_{N})\| \leq \epsilon \|u(\cdot)\|.$  
Otherwise we recompute 
the solution with $\kappa_{\mathrm{min}} \rightarrow \kappa_{\mathrm{min}}/2$ 
\footnote{This strategy proved useful in many experiments. 
  However we consider to refine it.}. Since for an anomalous 
mode the field is not damped at all we restrict the maximum $\xi_N$ to
$\xi_N<\pi/k_0/\epsilon.$ The pseudocode to the algorithm is given
in Algorithm~\ref{Algorithm:AdaptivePML}.
\begin{figure}
  \begin{center}
    \psfrag{Einc}[lc][lc][1.2][0]{$\VField{E}_{\mathrm{inc}}$}
    \psfrag{Eout}[lc][lc][1.2][0]{$\VField{E}_{\mathrm{out}}$}
    \psfrag{Eref}[lc][lc][1.2][0]{$\VField{E}_{\mathrm{ref}}$}
    \psfrag{x1}[lc][lc][1.2][0]{$x_1$}
    \psfrag{x3}[lc][lc][1.2][0]{$x_3$}
    \psfrag{theta}[lc][lc][1.2][0]{$\vartheta$}
    \includegraphics[width = 0.5\textwidth]{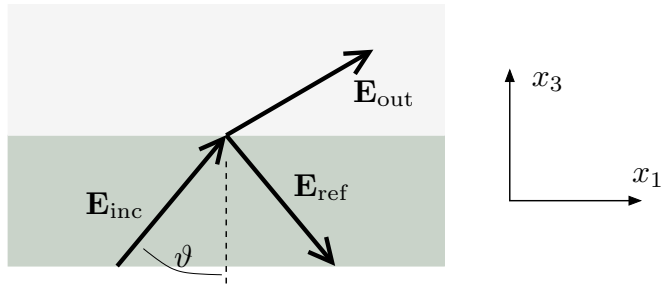}
    \caption{\label{Fig:TwoLayerScetch} Test problem for adaptive PML discretization. 
      A plane wave is incident under an angle $\vartheta$ from the lower material with 
      refractive index  $n_{\mathrm{sub}}=1.5$.
      The upper material consists of air ($n_{\mathrm{sup}}=1.0$). According to Snell's law
      the field is totally reflected for an incident angle greater or equal to the {\em critical
      angle} $\vartheta_{c}=180\cdot \mathrm{asin}(1.0/1.5)/\pi \approx 41.81.$
    }
  \end{center}
\end{figure}
\begin{figure}
  \begin{center}
    \includegraphics[width = 0.45\textwidth]{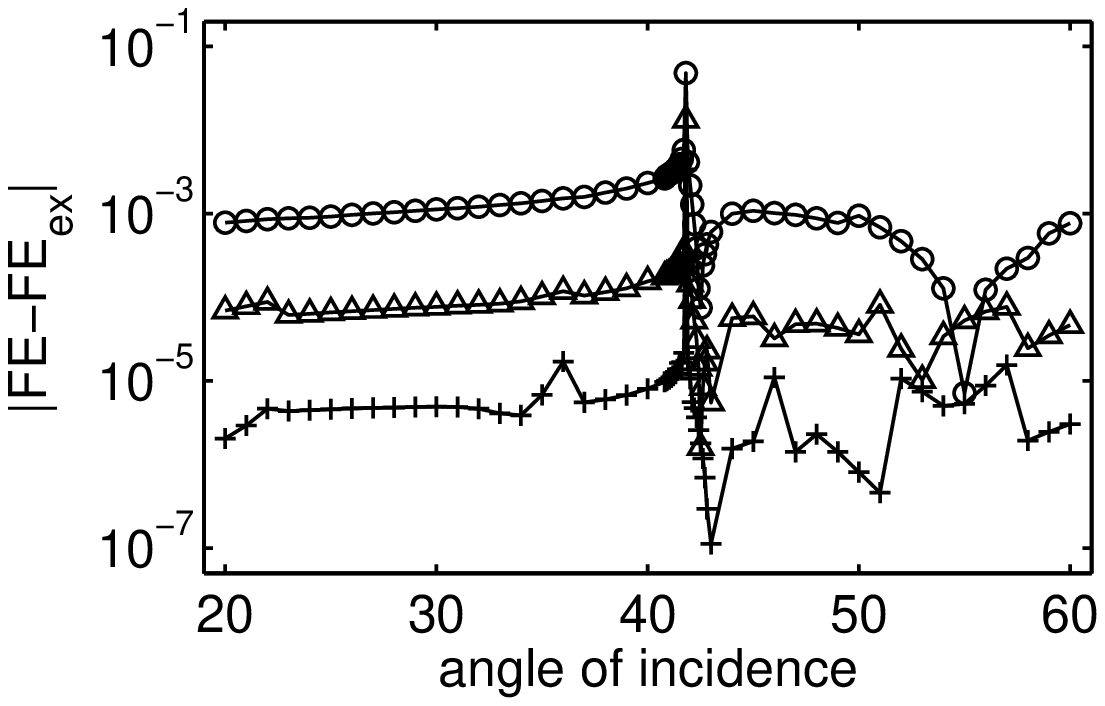}
    \includegraphics[width = 0.45\textwidth]{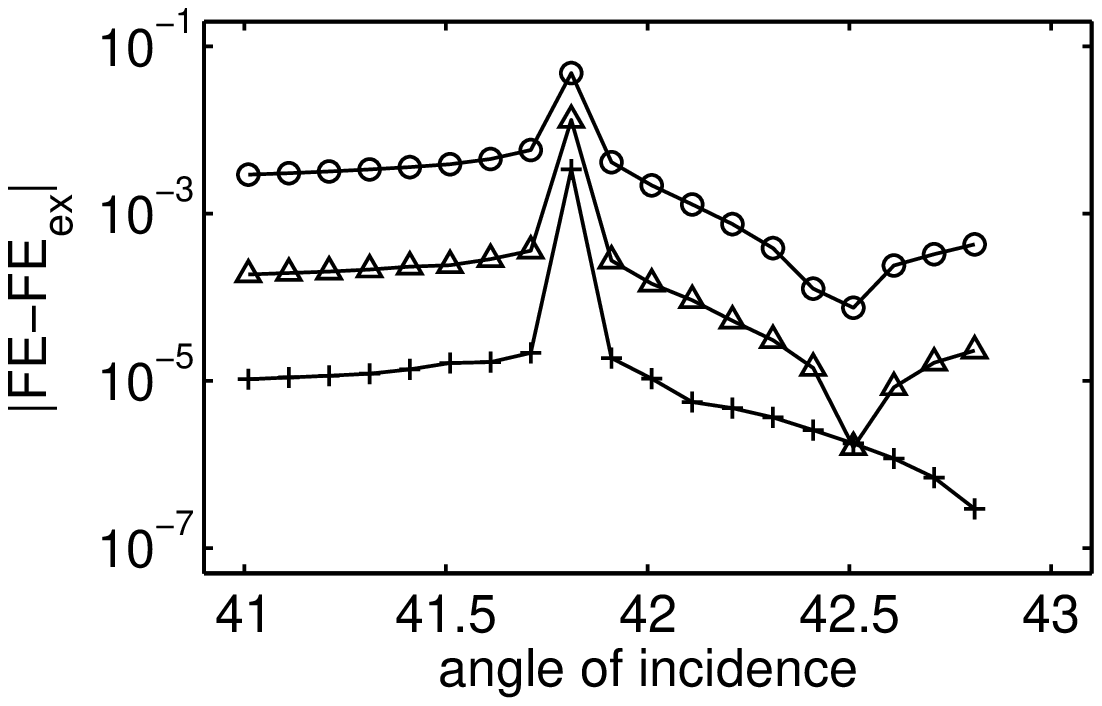}
    \caption{\label{Fig:PMLTestErrorScan} Left: 
      Field energy error in the interior domain. The three data sets ($\circ$, $\triangle$,
      +) correspond to different refinement levels of the interior domain.  Right: Zoom
      into the left figure near the critical angle.}
  \end{center}
\end{figure}

\begin{figure}
  \begin{center}
    \includegraphics[width = 0.45\textwidth]{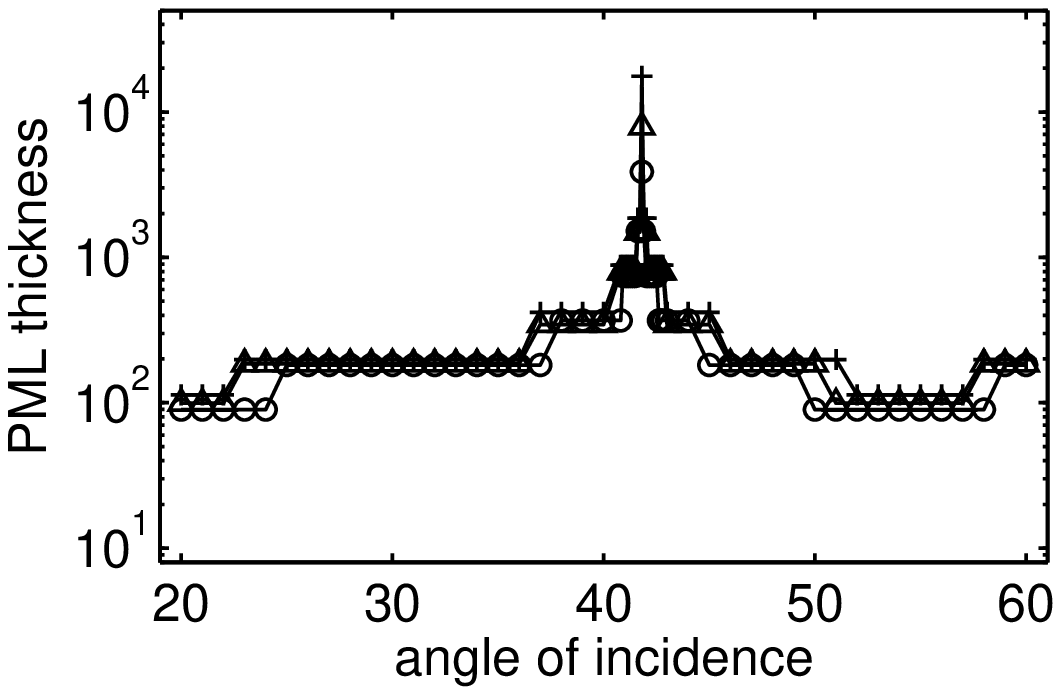}
    \includegraphics[width = 0.45\textwidth]{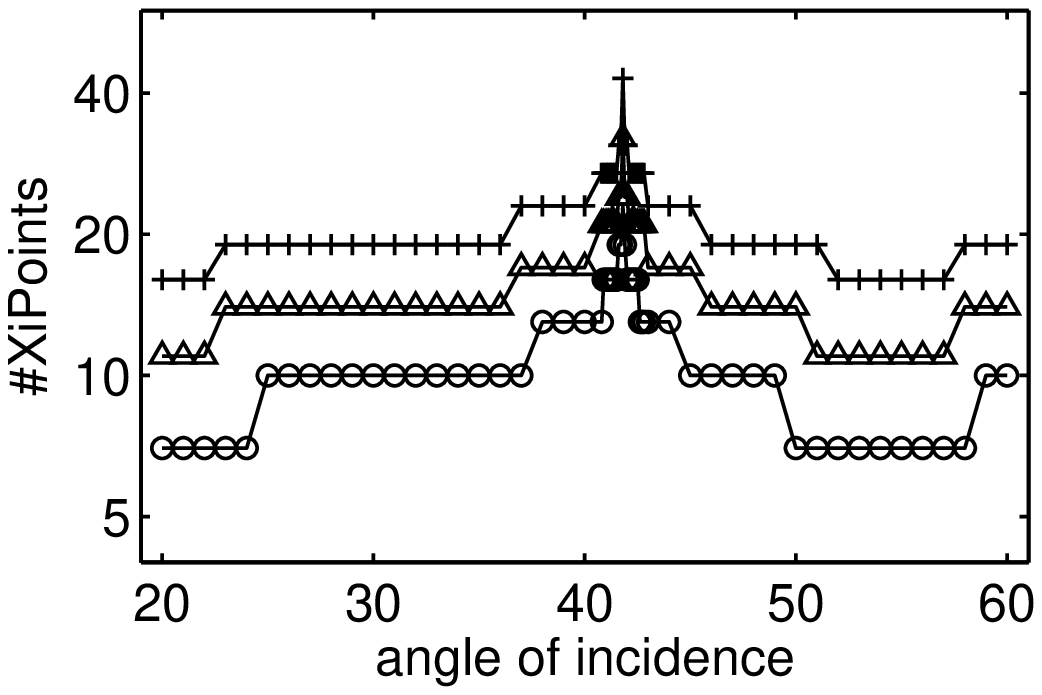}
    \caption{\label{Fig:PMLTestParameter} Left: 
      Thickness of the PML layer in unit lengths. At the critical angle the thickness is up to
      $10^{4}$ times larger than the size of the interior domain. Right: Number
      of discretization points $\xi_{j}$ used in the radial direction
      ($x_{2}$). Although the required thickness of the layer is huge the number of
      unknowns used in the PML layer remains moderate.}
  \end{center}
\end{figure}
\begin{table}
\begin{center}
\begin{tabular}{rrr}
Step & $\Delta E$ & $\Delta E'$  \\
\hline
 0 & 0.359850   &  0.335129  \\
 1 & 0.159358   &  0.166207  \\ 
 2 & 0.048779   &  0.049502  \\
 3 & 0.012911   &  0.012912  \\
 4 & 0.003274   &  0.003266  \\
 5 & 0.000205   &  0.000820  \\
 6 & 0.000206   &  0.000205  \\
 7 & 0.000051   &  0.000051 
\end{tabular}
\end{center}
\caption{Convergence of field energy at the critical angle of incidence. The first
column corresponds to the interior mesh refinement step. The relative error of
the electric field energy in the interior domain is given in the second
column, $\Delta E = |\|\VField{E}_{ex}\|_{L^{2}}^{2}-\|\VField{E}_{h}\|_{L^{2}}^{2}|/
\|\VField{E}_{ex}\|_{L^{2}}^{2}.$ 
The third column displays the relative error of the magnetic field energy 
$\Delta E' = |\|\curl \VField{E_{ex}}\|_{L^{2}}^{2}-\|\curl \VField{E_{h}}\|_{L^{2}}^{2}|/
\|\curl \VField{E_{ex}}\|$ is given. For fixed PML thickness the solution converges 
towards the analytical result as the interior mesh is refined.
\label{Tab:ConvergencAtWA}}
\end{table}
To demonstrate the performance of the adaptive PML algorithm we compute the
reflection of a plane wave at a material interface,
cf. Figure~\ref{Fig:TwoLayerScetch}. In $x_1-$direction we use Bloch periodic
boundary conditions \cite{Burger2005a}. We vary the angle of incidence from
$\vartheta=20^{\circ}$ to $\vartheta=60^{\circ}.$ Further the incoming field is
rotated along the $x_{3}$ axis by an angle of $45^{\circ},$ so that the incidence is
twofold oblique (conical). Hence the unit direction of the incoming field is equal
to $\hat{k} = (\cos 45^{\circ} \sin \vartheta, \cos \vartheta, \sin 45^{\circ} \sin
\vartheta).$ We use an interior domain of size $1.5 \times 1$ in wavelength
scales. To measure the error we compute the field energy within the interior domain
and compare it to the analytic value. In Figure~\ref{Fig:PMLTestErrorScan} the
error is plotted for different refinement levels of the interior domain. The ``+''
line corresponds to the finest level. In Figure~\ref{Fig:PMLTestParameter} the
automatically adapted thickness of the PML is plotted (left) and the number of
discretization points $N$ in $\xi$ direction (right). As expected a huge layer is
constructed automatically at the critical angle, 
whereas the total number of discretization points remains
moderate. As can be seen in Figure~\ref{Fig:PMLTestErrorScan} the maximum error
appears at the critical angle. From that one may suspect a failure of the automatic
PML adaption. But a closer analysis reveals that the chosen discretization in the PML
layer is sufficient as can be seen from Table~\ref{Tab:ConvergencAtWA}. Here the thickness
of the perfectly matched layer has been fixed and we further refined
the interior domain. 
By this means we observe convergence to the true solution but the
convergence rate is halved at the critical angle. Hence the maximum error at the
critical angle is caused by an insufficient interior discretization. We conjecture that
this is due to a dispersion effect. Since near the critical angle 
the wave $\VField{E}_{\mathrm{out}}$ is traveling mainly along the $x_{1}-$
direction it reenters the periodic domain, leading to large ``path length''.

\section{VARIATIONAL FORMULATION}
\label{weak_section}
So far the overall scattering problem was given as an interior domain problem coupled to an
exterior domain problem via boundary matching conditions.
In this section we give (without proof) a variational problem in $H(\curl, \Omega \cup \Omega_{\mathrm{ext}})$
for the computation of the composed field $\tilde{\VField{E}}$ with $\tilde{\VField{E}}=\VField{E}_{\mathrm{int}}$ in $\Omega_{\mathrm{int}}$ and 
$\tilde{\VField{E}}=\VField{E}_{\mathrm{sc, \gamma}}+\Pi({\VField{E}}_{\mathrm{inc}}\times \vec{n})$ in $\Omega_{\mathrm{ext}}$. Details for the 2D case are given in our paper~\cite{Zschiedrich03a}. Here $\Pi$ is the extension operator defined as
$\Pi({\VField{E}}_{\mathrm{inc}}\times \vec{n}) = \chi_{[0, \epsilon)}(1-\xi/\epsilon)({\VField{E}}_{\mathrm{inc}}\times \vec{n}).$

For each face $F$ of the transparent boundary $J_{F}(\eta_{1}, \eta_{2}, \xi)$ denotes 
the Jacobian of the mapping $Q_{F}(\eta_{1}, \eta_{2}, \xi).$ Further we
introduce the {\em pulled back} field $\VField{u}_{*}(\eta_{1}, \eta_{2}, \xi) = 
J^{\mathrm{t}}\VField{u}(Q_{F}(\eta_{1}, \eta_{2}, \xi))$ for any field defined on $\Omega_{\mathrm{ext}}.$
With the definition
\[
\curl_{\gamma} = 
(
\partial_{\eta_{2}}-\frac{1}{\gamma}\partial_{\xi}, 
-\partial_{\eta_{1}}+\frac{1}{\gamma}\partial_{\xi},
\partial_{\eta_{1}}-\partial_{\eta_{2}}) 
\]
and the transformed tensors $ \mu_{*} = |\mathrm{J}| \mathrm{J}^{-1} \mu \mathrm{J}^{-\mathrm{t}}$ and  $ \varepsilon_{*} =  |\mathrm{J}| \mathrm{J}^{-1} \varepsilon \mathrm{J}^{-\mathrm{t}}$ the composed field 
$\tilde{\VField{E}}$ satisfies
\begin{eqnarray*}
\int_{\Omega_{\mathrm{int}}} \curl \VField{\Psi}  \mu^{-1} \curl \tilde{\VField{E}}-
\omega^{2} \VField{\Psi}\varepsilon \tilde{\VField{E}}+
\gamma \sum_{F} \int_{P_{u}}  \curl_{\gamma} \VField{\Psi}_{*}  \mu^{-1}_{*} \curl_{\gamma} \tilde{\VField{E}}_{*}-
\omega^{2} \VField{\Psi}_{*}\varepsilon_{*} \tilde{\VField{E}}_{*} & = & \\
-\int_{\partial_{\Omega_{\mathrm{int}}}}  
\VField{\Phi} \cdot \mu^{-1} \curl_{3}
\VField{E}_{\mathrm{inc}} \times \vec{n} +
\gamma \sum_{F} \int_{P_{u}}  \curl_{\gamma} \VField{\Psi}_{*}  \mu^{-1}_{*} \curl_{\gamma} \Pi ({\VField{E}}_{\mathrm{inc}}\times \vec{n})-\omega^{2} \VField{\Psi}_{*}  \varepsilon \Pi ({\VField{E}}_{\mathrm{inc}}\times \vec{n}).
\end{eqnarray*}
Although this equation looks complicated it can be easily discretized with finite elements. In fact the terms in the sum over the faces $F$ are already given in unit coordinates of the prism. So it is advantageous to use a prismatoidal mesh in the exterior domain. This way we fix the {\em global} discretization points $\{\xi_{0} = 0, \xi_{1}, \cdots, \xi_{N}\}$ as described in the previous section and split the truncated exterior domain $\Omega_{\xi_{N}}$ into the prisms $Q_{F}(P_{\xi_{i+1}} \setminus P_{\xi_{i}})$ with $i<N.$ 
In the interior domain we use a tetrahedral mesh which we fit non-overlapping to the exterior domain mesh. Introducing the bilinear forms
\begin{eqnarray*}
a_{\mathrm{int}}\left(\VField{\Psi}, \VField{\Phi}\right) & = & \int_{\Omega_{\mathrm{int}}} \curl \VField{\Psi}  \mu^{-1} \curl \tilde{\VField{E}} \\
b_{\mathrm{int}}\left(\VField{\Psi}, \VField{\Phi}\right) & = & \int_{\Omega_{\mathrm{int}}} \VField{\Psi}\varepsilon \tilde{\VField{E}} \\
a_{\gamma}\left(\VField{\Psi}, \VField{\Phi}\right) & = & \gamma \sum_{F} \int_{P_{\rho}}  \curl_{\gamma} \VField{\Psi}_{*}  \mu^{-1}_{*} \curl_{\gamma} \tilde{\VField{E}}_{*} \\
b_{\gamma}\left(\VField{\Psi}, \VField{\Phi}\right) & = & \gamma \sum_{F} \int_{P_{\rho}} 
\VField{\Psi}_{*}\varepsilon_{*} \tilde{\VField{E}}_{*} \\
a\left(\VField{\Psi}, \VField{\Phi}\right) & = & a_{\mathrm{int}}\left(\VField{\Psi}, \VField{\Phi}\right) + a_{\gamma}\left(\VField{\Psi}, \VField{\Phi}\right) \\
b\left(\VField{\Psi}, \VField{\Phi}\right) & = & b_{\mathrm{int}}\left(\VField{\Psi}, \VField{\Phi}\right) + b_{\gamma}\left(\VField{\Psi}, \VField{\Phi}\right)
\end{eqnarray*}
and 
\begin{equation*}
g\left(\Psi\right) = -\int_{\partial_{\Omega_{\mathrm{int}}}}  
\VField{\Phi} \cdot \mu^{-1} \curl_{3}
\VField{E}_{\mathrm{inc}} \times \vec{n}
\end{equation*}
the variational problem truncated to $\Omega_{\mathrm{int}} \cup \Omega_{\rho}$ can be casted to
\[
a\left(\VField{\Psi}, \tilde{\VField{E}}\right)- 
\omega^{2}
b\left(\VField{\Psi}, \tilde{\VField{E}}\right) = 
g\left(\Psi\right)+
a_{\gamma}\left(\VField{\Psi},  \Pi ({\VField{E}}_{\mathrm{inc}}\times \vec{n})\right)-
\omega^{2}
b_{\gamma}\left(\VField{\Psi},  \Pi ({\VField{E}}_{\mathrm{inc}}\times \vec{n})\right) 
\]
for all $\VField{\Psi} \in H(\curl, \Omega_{\mathrm{int}} \cup \Omega_{\rho}).$ 
To discretize this variational problem we use  Nedelec's vectorial finite elements~\cite{Monk2003a} with local ansatz functions $\{\VField{v}_{1}, \VField{v}_{2}, \dots, \VField{v}_{n} \}.$ Making the ansatz 
$\tilde{\VField{E}}=\sum u_{i}\VField{v}_{i}$ this yields the algebraic system
\[
(A-\omega^{2})Bu = f
\]
with $f_{i}=
g\left(\VField{v_{i}} \right)+
a_{\gamma}\left(\VField{v_{i}}, \Pi ({\VField{E}}_{\mathrm{inc}}\times \vec{n})\right)-
\omega^{2}
b_{\gamma}\left(\VField{v_{i}}, \Pi ({\VField{E}}_{\mathrm{inc}}\times \vec{n})\right),$ 
$A_{i, j} = a\left(\VField{v_{i}}, \VField{v_{j}}\right)$ and $B_{i,j}$ accordingly.

\section{TIME DOMAIN PRECONDITIONER}
\label{td_section}
In this section we propose a novel preconditioner for the algebraic system
\begin{equation}
\label{Eqn:THDISCRETE}
(A-\omega^{2}B)u = f
\end{equation}
derived in the last section.
Since $\omega^{2}>0$ this system is indefinite. Hence standard multigrid methods will suffer from
slow convergence rates or may even not converge. Other numerical methods like the Finite Difference Time Domain method do not start from the time-harmonic Maxwell's equations. 
Instead they simulate temporal transient effects. 
For practical purposes the computation time is prohibitively large until the steady state
is reached~\cite{Burger_benchmark}. 
Even worse, the usage of an explicite time stepping scheme forces the usage of very small time steps to avoid instabilities.
  
Here we propose a preconditioner for the time-harmonic system which makes use 
of the fact that the solution we want to compute is the steady state solution to a transient process. This is
the reason why we call this preconditioner ``time domain'' preconditioner. Instead of using time dependent Maxwell's equations one may use another dynamical system whose steady state solution is the field in mind.
For example the above solution $u$ is the steady state solution to the time dependent problem
\[
i \frac{\mathrm{d}}{\mathrm{d}t} B u(t)= A u(t) - fe^{-i\omega^{2}t}
\]
which looks like the time dependent Schr\"odinger equation.
But we may also start from a time discrete system, such as
\begin{equation}
\label{Eqn:IMPEULER}
i\frac{1}{\tau}(Bu_{n+1}-Bu_{n})=Au_{n+1} - p^{n+1}f 
\end{equation}
with $p=(\frac{1}{1+i\omega^{2}\tau})$ corresponding to a time discretization of the above Schr\"odinger like equation with the implicite Euler method. 
One proves that if the original system has a steady state solution
then the sequence $\{u_{0}, p^{-1}u_{1}, p^{-2}u_{2}, \cdots\}$ converges to the solution 
$u$ of Equation~\eqref{Eqn:THDISCRETE} indepently of the selected time step $\tau.$ In our code
we typically fix $\tau = 0.1/\omega^{2}.$ Starting from a randomized initial guess $u_{0}$ we compute a fixed number $N$ of iterations to the recursion formula~\eqref{Eqn:IMPEULER} which yields
the sequence $\{u_{0}, p^{-1}u_{1}, \dots, p^{-N}u_{N}\}.$ In each iteration step the arising system is solved by an multigrid method up to a moderate accuracy. We then compute the minimum residual solution within the space spanned by the last $M<N$ vectors in this sequence.
 The so constructed approximate solver is used as a preconditioner for a standard iterative method for indefinite problems such as GMRES or BCGSTAB~\cite{Freund:92a}. 

Other discrete schemes may be used to improve the convergence to the steady state solution.
For example it is promising to use schemes stemming from higher
order Runge-Kutta methods or multi-step methods for the discretization 
of the original wave equation or the Schr\"odinger like equation above. 

\section{RESONANCE PROBLEMS}
A resonance is a purely outgoing field which satisfies the time harmonic Maxwell's equation
for a certain $\omega \in \cnum.$ 
We again assume that an expansion as in Equation~\eqref{Eqn:PMLExpansion} is valid but
we must drop the assumption $\Im k_{\xi}(\alpha)\geq 0$. Hence a resonance mode may 
exponentially grow with the generalized distance $\xi.$ In this case we must choose
$\sigma$ large enough in the PML method to achieve an exponential damping of the complex 
continued solution. Using a finite element discretization as in the previous section we end up with the algebraic eigenvalue problem
\begin{equation}
\label{Eqn:EVP}
Au=\omega^{2}Bu.
\end{equation}
  
\section{META-MATERIALS: SPLIT RING RESONATORS}
\label{srr_section}

Split-ring resonators (SRR's) can be understood as small $LC$ circuits 
consisting of an inductance $L$ and a capacitance $C$.
The circuit can be driven by applying external electromagnetic 
fields.
Near the resonance frequency of the $LC-$oscillator
the induced current can lead to a magnetic field opposing the external 
magnetic field. 
When the SRR's are small enough and closely packed -- such that the system 
can be described as an effective medium -- 
the induced opposing magnetic field corresponds to an effective 
negative permeability, $\mu<0$, of the medium.

Arrays of gold SRR's with resonances in the NIR and in the optical 
regime can be experimentally realized using electron-beam lithography, 
where typical dimensions are one order of magnitude smaller than 
NIR wavelengths. 
Details on the production can be found in Linden and Enkrich~\cite{Linden2004a,Enkrich2005a}.

\begin{figure}
  \begin{center}
    \psfrag{(a)}[lc][lc][1.2][0]{(a)}
    \psfrag{(b)}[lc][lc][1.2][0]{(b)}
    \includegraphics[width = 1.0\textwidth]{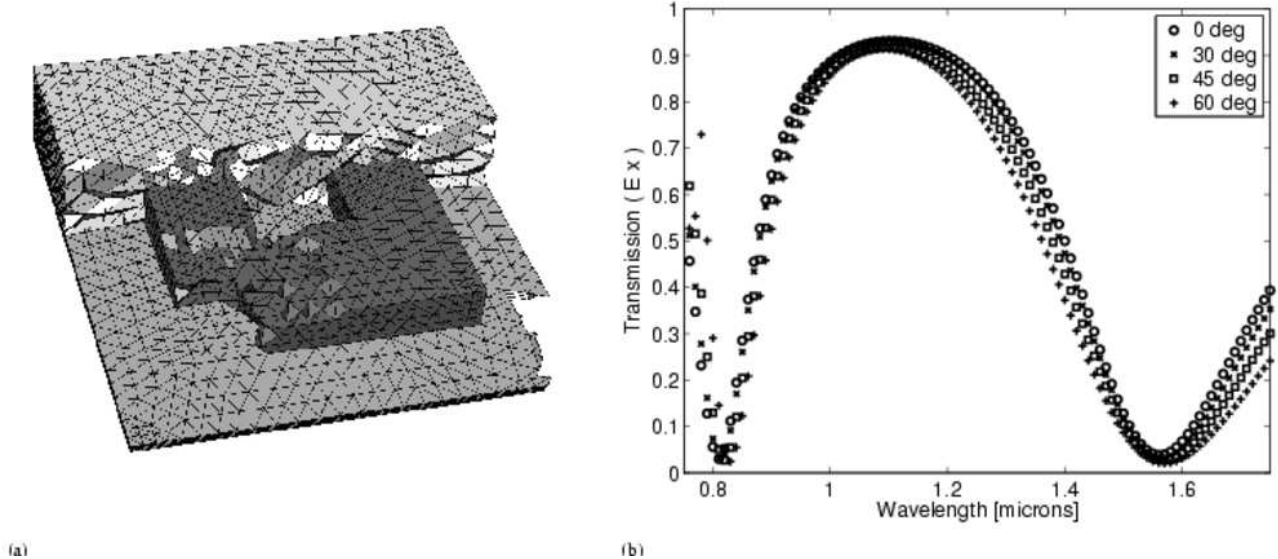}
    \caption{
    \label{srr_1}     
    (a) Visualization of a part of the tetrahedra of the spatial discretization 
    of the interior SRR geometry. Dark gray tetrahedra: gold SRR; light 
    gray: air; gray: ITO. Periodic boundary conditions apply at the right/left and 
    front/back. Prism elements discretizing the exterior domain (on the top/bottom)
    are not shown. 
    (b) Transmission spectra of light fields incident onto an SRR for different 
    angles of incidence (For details see also~\cite{Burger2005w}). 
    The transmission minimum at $\lambda \sim 1.5\mu$m is due to 
    the excitation of the fundamental resonance of the SRRs.  
    {\footnotesize (See original publication for images with higher resolution.)}
     }
  \end{center}
\end{figure}

Due to the small dimensions of the $LC$ circuits their resonances are 
in the NIR and optical regime~\cite{Enkrich2005a}.
Figure~\ref{srr_1}(a) shows the tetrahedral discretization of the 
interior domain of the geometry. 
Figure~\ref{srr_1}(b) shows results from FEM simulations of light scattering 
off a periodic array of SRR's for different angles of the incident 
light~\cite{Burger2005w}. 
At  $\lambda \sim 1.5\mu$m the transmission is strongly reduced due to 
the excitation of a resonance of the array of SRR's.
This excitation occurs for all investigated angles of incidence.

When one is interested to learn about resonances, 
obviously it is rather indirect and time-consuming to calculate the scattering 
response of some incident light field and then to conclude the properties of 
the resonance. 
We have therefore also directly computed resonances of SRR's by 
solving Eqn.~\eqref{Eqn:EVP}.
Special care has to be taken in constructing appropriate PML layers, as in 
this case the previously described adaptive strategy for the PML is more involved. 
We have therefore set these parameters by hand. We have computed the fundamental resonance
with $\omega=1.302023 \cdot 10^{15} - 0.399851 \cdot 10^{15}i.$

\section{PYRAMIDAL NANO-RESONATOR}
\label{pyramid_section}
\begin{figure}
  \begin{center}  
    \psfrag{x}[lc][lc][1.2][0]{x}
    \psfrag{y}[lc][lc][1.2][0]{y}
    \psfrag{z}[lc][lc][1.2][0]{z}
  \includegraphics[width = 1.0\textwidth]{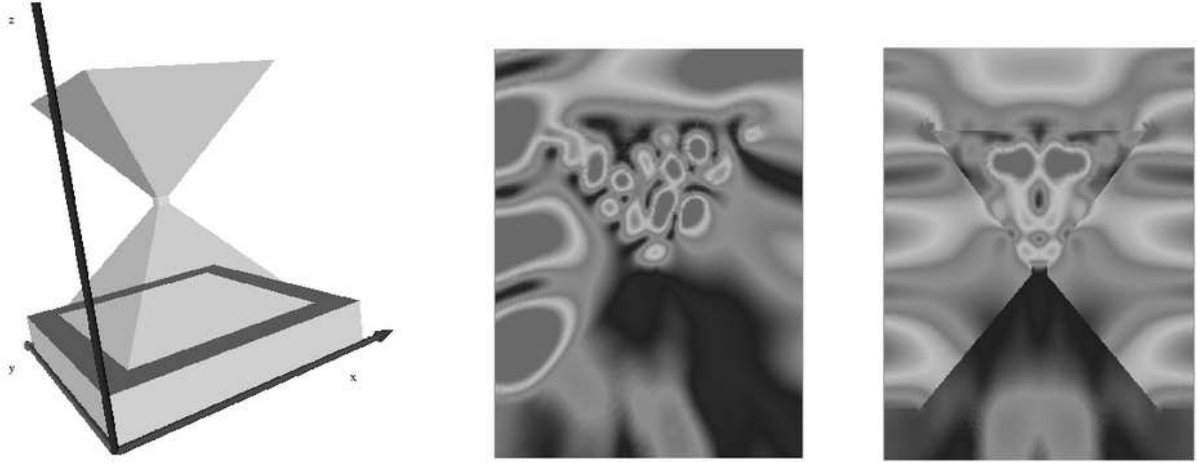}  
    \caption{
    \label{Fig:Pyramide} Pyramidal nano resonator. The structure in mounted
on a Gallium Arsenid (GaAs) substrate. The middle figure shows the field amplitude in the $x$-$z$ plane. The right figure shows the field amplitude in the $y$-$z$ plane.
    {\footnotesize (See original publication for images with higher resolution.)}
    }  
 \end{center}
\end{figure}

This type of nano resonators is proposed to be used as optical element in quantum information processing~\cite{Loeffler}. The structure is as depicted in Figure~\ref{Fig:Pyramide} (left). We have simulated the illumination of the structure with a plane wave of a vacuum wavelength $\lambda_{0}=1.55 \mu m$ and a unit direction $\hat{k}=(\sqrt{0.5}, 0, -\sqrt{0.5}).$ The incident field
was polarized in $x$-direction. Figure~\ref{Fig:Pyramide} shows the field amplitude 
in the computational domain. More than five million of unknowns were used in the discretization. With the preoconditioner proposed in Section~\ref{td_section} (N=20) the GMRES method exhibited a convergence rate of $0.8.$

\section*{ACKNOWLEDGMENTS}       
We thank R. Klose, A. Sch\"adle, P. Deuflhard and R. M{\"a}rz for fruitful 
discussions, and we acknowledge support by the initiative DFG Research 
Center {\em Matheon} of the Deutsche Forschungsgemeinschaft, DFG, and 
by the DFG under contract no. BU-1859/1.
\bibliography{lit1,lit2,other}   

\begin{thebibliography}{10}

\bibitem{Linden2004a}
S.~Linden, C.~Enkrich, M.~Wegener, C.~Zhou, T.~Koschny, and C.~Soukoulis,
  ``Magnetic response of metamaterials at 100 {T}erahertz,'' {\em Science}~{\bf
  306}, p.~1351, 2004.

\bibitem{Enkrich2005a}
C.~Enkrich, M.~Wegener, S.~Linden, S.~Burger, L.~Zschiedrich, F.~Schmidt,
  C.~Zhou, T.~Koschny, and C.~M. Soukoulis, ``Magnetic metamaterials at
  telecommunication and visible frequencies,'' {\em Phys. Rev. Lett.}~{\bf 95},
  p.~203901, 2005.

\bibitem{Veselago1968a}
V.~G. Veselago, ``The electrodynamics of substances with simultaneously
  negative values of $\epsilon$ and $\mu$,'' {\em Sov. Phys. Usp.}~{\bf 10},
  p.~509, 1968.

\bibitem{Pendry2000a}
J.~B. Pendry, ``Negative refraction makes a perfect lens,'' {\em Phys. Rev.
  Lett.}~{\bf 85}, p.~3966, 2000.

\bibitem{Berenger94}
J.-P. B\'erenger, ``A perfectly matched layer for the absorption of
  electromagnetic waves,'' {\em J. Comput. Phys.}~{\bf 114}(2), pp.~185--200,
  1994.

\bibitem{Schmidt2002a}
F.~Schmidt, ``{A New Approach to Coupled Interior-Exterior Helmholtz-Type
  Problems: Theory and Algorithms},'' 2002.
\newblock Habilitation thesis, Freie Universitaet Berlin.

\bibitem{Lassas:98a}
M.~Lassas and E.~Somersalo, ``On the existence and convergence of the solution
  of {PML} equations.,'' {\em Computing No.3, 229-241}~{\bf 60}(3),
  pp.~229--241, 1998.

\bibitem{Lassas:01a}
M.~Lassas and E.~Somersalo, ``Analysis of the {PML} equations in general convex
  geometry,'' in {\em Proc. Roy. Soc. Edinburgh Sect. A 131},  (5),
  pp.~1183--1207, 2001.

\bibitem{Hohage01b}
T.~Hohage, F.~Schmidt, and L.~Zschiedrich, ``Solving time-harmonic scattering
  problems based on the pole condition:{C}onvergence of the {PML} method,''
  Tech. Rep. ZR-01-23, Konrad-Zuse-Zentrum (ZIB), 2001.

\bibitem{Schaedle:06a}
A.~Sch{\"a}dle, L.~Zschiedrich, S.~Burger, R.~Klose, and F.~Schmidt, ``{D}omain
  {D}ecomposition {M}ethod for {M}axwell's {E}quations: {S}cattering off
  {P}eriodic {S}tructures,'' {\em in preparation} , 2006.

\bibitem{Petit1980a}
R.~Petit, {\em Electromagnetic Theory of Gratings}, Springer-Verlag, 1980.

\bibitem{Burger2005a}
S.~Burger, R.~Klose, A.~Sch\"adle, F.~Schmidt, and L.~Zschiedrich, ``{FEM}
  modelling of 3d photonic crystals and photonic crystal waveguides,'' in {\em
  Integrated Optics: Devices, Materials, and Technologies IX},  Y.~Sidorin and
  C.~A. W\"achter, eds., ~{\bf 5728}, pp.~164--173, Proc. SPIE, 2005.

\bibitem{Zschiedrich03a}
L.~Zschiedrich, R.~Klose, A.~Sch\"adle, and F.~Schmidt, ``A new finite element
  realization of the {P}erfectly {M}atched {L}ayer {M}ethod for {H}elmholtz
  scattering problems on polygonal domains in 2{D},'' {\em J. Comput Appl.
  Math.} , 2005.
\newblock in print; published online.

\bibitem{Monk2003a}
P.~Monk, {\em Finite Element Methods for {M}axwell's Equations}, Claredon
  Press, Oxford, 2003.

\bibitem{Burger_benchmark}
S.~Burger, R.~K{\"o}hle, L.~Zschiedrich, W.~Gao, F.~Schmidt, R.~März, and
  C.~N{\"o}lscher, ``Benchmark of {FEM}, {W}aveguide and {FDTD} {A}lgorithms
  for {R}igorous {M}ask {S}imulation,'' in {\em Photomask Technology},  J.~T.
  Weed and P.~M. Martin, eds., ~{\bf 5992}, pp.~368--379, SPIE.

\bibitem{Freund:92a}
R.~Freund, G.~Golub, and N.~Nachtigal, ``Iterative solution of linear
  systems,'' {\em Acta Numerica} , 1992.

\bibitem{Burger2005w}
S.~Burger, L.~Zschiedrich, R.~Klose, A.~Sch\"adle, F.~Schmidt, C.~Enkrich,
  S.~Linden, M.~Wegener, and C.~M. Soukoulis, ``Numerical investigation of
  light scattering off split-ring resonators,'' in {\em Metamaterials},
  T.~Szoplik, E.~{\"O}zbay, C.~M. Soukoulis, and N.~I. Zheludev, eds., ~{\bf
  5955}, pp.~18--26, Proc. SPIE, 2005.

\bibitem{Loeffler}
H.~K. W.~L\"offler, ``private communication, {CFN} {K}arlsruhe,'' 2005.

\end{thebibliography}
\bibliographystyle{spiebib}  
\end{document}